# Enhancement of Entanglement via Incoherent Collisions

Xihua Yang[1*], Mingfei Cheng[1], and Min Xiao[2,3+]

[1]*Department of Physics, Shanghai University, Shanghai 200444, China*

[2]*Department of Physics, University of Arkansas, Fayetteville, Arkansas 72701, USA*

[3]*National Laboratory of Solid State Microstructures and School of Physics, Nanjing University, Nanjing 210093, China*

(Dated: July 16, 2023)

In contrast to the general thought that the collisions are intrinsically dephasing in nature and detrimental to quantum entanglement at room or higher temperatures, here, we show that in the conventional ladder-type three-level electromagnetically induced transparency (EIT) configuration, when the probe field intensity is not very weak as compared to the pump field, the entanglement between the bright pump and probe fields can be remarkably enhanced with the increase of the collisional decay rates in a moderate range in an inhomogeneously-broadened atomic system. The strengthened entanglement results from the enhancement of constructive interference and suppression of destructive interference between one-photon and multi-photon transition pathways. Our results clearly indicate that the collisions offer a promising alternative to enhance entanglement at room or higher temperatures despite of the dephasing nature, which provides great convenience for experimental implementation, and opens new prospects and applications in realistic quantum computation and quantum information processing.

It is generally thought that the collisions are dephasing in nature and detrimental to quantum coherence and interference as well as to squeezing and entanglement at room or higher temperatures. However, the collision-induced quantum effects have been extensively studied. The pressure-induced extra resonance resulted from collision-aided quantum interference was investigated first by Bloembergen in four-wave mixing (FWM) [1, 2] and then by Grynberg in nonlinear spectroscopy [3]. The quantum interference between two collision-assisted excitation pathways in both the frequency and time domains has been demonstrated experimentally and theoretically [4-6]. It was pointed out that the nature of quantum interference in the various optically-driven three-level systems critically depends on the excitation scheme and the dephasing collisions can even change the nature [7, 8]. Moreover, the control of coherent population transfer can be achieved via the collision-assisted EIT and electromagnetically induced absorption (EIA) in a closed inverse-Y-type four-level system with the stimulated Raman adiabatic passage technique [9].

On the other hand, the generation of squeezing and entanglement only with the existence of suitable dephasing rates has also been investigated. It was shown that the realization of electromagnetically induced entanglement (EIE) [10] and electromagnetically induced squeezing (EIS) of atomic spin [11] would rely on suitable coherence decay rate of the lower doublet in the traditional $\Lambda$-type three-level EIT configuration, and no squeezing or entanglement would exist with zero dephasing rate. This counterintuitive behavior has been observed as well for the generation of pump-probe intensity correlation [12] and squeezed or entangled states of light [13, 14] in the $\Lambda$-type coherent population trapping or EIT configuration.

Motivated by the EIA via incoherent collisions studied in Ref. [8], we present a convenient and efficient way to enhance the bipartite entanglement between the bright pump and probe fields via incoherent collisions in the inhomogeneously-broadened ladder-type atomic system. We show that when the probe field intensity is not very weak as compared to the pump field, the degree of the bipartite entanglement can be dramatically enhanced with the increase of the collisional decay rates in a moderate range due to the collision-assisted one-photon and multi-photon quantum interference. This method would greatly facilitate the generation and enhancement of bipartite entanglement between bright light fields at room or higher temperatures, and may find broad and potential applications in practical quantum computation, quantum communication, and quantum networks.

The ladder-type three-level atomic system driven by a strong coherent pump field and a relatively weak probe field denoted by the quantum operators $a_2$ and $a_1$, is shown in Fig. 1a, where the levels 1, 2, and 3 correspond, respectively, to the levels 5S (F=3), 5P$_{3/2}$, and 5D$_{5/2}$ of the $^{85}$Rb atom. The probe field with frequency $\omega_1$ and pump field with frequency $\omega_2$ couple the levels 1 and 2 and levels 2 and 3 with the frequency detunings $\Delta_1 = \omega_1 - \omega_{21}$ and $\Delta_2 = \omega_2 - \omega_{32}$, respectively. We denote $2\gamma_1$ and $2\gamma_2$ as the population decay rates from level 2 to level 1 and level 3 to level 2, respectively, and $\gamma_{ij}$ (i ≠ j) as the atomic coherence decay rate between levels i and j. Apart from the radiative relaxation, the atoms also undergo collisions, and the collision-induced coherence decay rate between levels i and j is denoted by $\gamma_{ijp}$ (i ≠ j). In what follows, we take into account the quantum features of both the pump and probe fields, and examine the bipartite entanglement between the two bright light fields in the inhomogeneously-broadened ladder-type atomic system under different collisional damping rates.

As is well known, the successful generation of entanglement using initially coherent light fields in an atomic system critically relies on the strong nonlinear interaction (e.g. FWM process) of light fields with atoms. In fact, FWM has proven to be an efficient process to produce entanglement, as demonstrated by the generation of entangled Stokes and anti-Stokes photons in the Λ-type atomic system [15-22]. With this in mind, and based on the collision-induced EIA in Ref. [8], we try to test the bipartite entanglement between the pump and probe fields with comparable intensity under different collisional decay rates shown in Fig. 1a. Figure 1b schematically displays the one-photon and multi-photon absorption up to the fifth-order terms of the probe field. With the similar analysis to that in Refs. [8, 23, 24], considering the two-step two-photon excitation (TSTPE) as the dominant contribution to the probe absorption for the higher-order terms of the probe field, we solve the Heisenberg-Langevin equations iteratively to the fifth-order of the mean value of the collective atomic operators $\langle \sigma_{21} \rangle$, and get the expression for the probe absorption coefficient with the consideration of the Doppler broadening to be

$$\alpha(\Delta_1) \propto \frac{\gamma_{21}}{g_1 \langle a_1 \rangle} \int dv D(v) \operatorname{Im} \langle \sigma_{21} \rangle = \int dv D(v) \{ \frac{\gamma_{12}^2}{\gamma_{12}^2 + \Delta_1^2} - \operatorname{Re}[\frac{\gamma_{12}(g_2 \langle a_2 \rangle)^2}{(\gamma_{12} + j\Delta_1)^2 (\gamma_{13} + j(\Delta_1 + \Delta_2))}]$$
$$+ \frac{(g_1 \langle a_1 \rangle)^2 (g_2 \langle a_2 \rangle)^2}{4 \gamma_1 \gamma_2 \gamma_{12} \gamma_{23}} \frac{\gamma_{23}^2}{\gamma_{23}^2 + \Delta_2^2} (\frac{\gamma_{12}^2}{\gamma_{12}^2 + \Delta_1^2})^2 \}, \tag{1}$$

where $g_{1(2)} = \mu_{12(23)} \cdot \varepsilon_{1(2)} / \hbar$ is the atom-field coupling constant with $\mu_{12(23)}$ being the dipole moment for the 1-2 (2-3) transition and $\varepsilon_{1(2)} = \sqrt{\hbar \omega_{1(2)} / 2\epsilon_0 V}$ being the electric field of a single probe (pump) photon, $\epsilon_0$ is the free space permittivity and $V$ is the interaction volume with length $L$ and beam radius $r$, and $D(v) = \exp(-v^2/\mu^2)/(\sqrt{\pi}\mu)$ is the normalized Doppler distribution with $\mu$ being the root-mean-square atomic velocity. Obviously, Eq. (1) is essentially the same as that in Refs. [8, 23, 24], only with $g_1 \langle a_1 \rangle$ and $g_2 \langle a_2 \rangle$ replaced by the Rabi frequencies of the probe and pump fields treated in the semi-classical density-matrix approach.

The entanglement feature of the pump and probe fields with comparable intensity under different collisional dephasing rates can be intuitively understood in terms of the nonlinear interaction between the laser fields and atomic medium shown in Fig. 1b. The one-photon and multi-photon excitation processes presented in Fig. 1b can be well described by the three components in Eq. (1) for the probe absorption. The first term comes from the traditional one-photon linear absorption; the second is from the lowest-order term in the pump field, which is the origin of EIT; and the third term consists of a five-photon process, representing the TSTPE. Clearly, both of the nonlinear processes EIT and TSTPE have contributions to the generation of the entanglement between the pump and probe fields. As discussed in Ref. [10], the EIT process is essentially a FWM process, which can be equivalently regarded as a closed-loop light-atom interaction, and in the present scheme, the FWM process involves the absorption of one probe photon and one pump photon and subsequent emission of one pump photon and one probe photon. Since every probe photon absorption (emission) is always accompanied by absorbing (emitting) one pump photon, strong quantum correlation and bipartite entanglement between the pump and probe fields can be produced. In the same way, the TSTPE process is essentially a six-wave mixing (SWM) process, involving the simultaneous absorption (emission) of one probe photon and one pump photon and the absorption of another probe photon accompanied by the emission of a further probe photon, which can also lead to strong quantum correlation and bipartite entanglement between the pump and probe fields.

However, as seen from Eq. (1), the second term has a minus sign with respect to the first term, and the destructive interference between the one-photon and three-photon transition pathways results in EIT, which would trap atoms in the level 1; subsequently, the FWM process as well as the bipartite entanglement would be weakened due to the EIT-induced reduction of population transfer, and even no entanglement would exist if the dephasing rate $\gamma_{13}$ were zero, as evidenced in the $\Lambda$-type three-level atomic system in Ref. [10]. On the other hand, the TSTPE term has the same sign as the first term, and the constructive interference between the one-photon and five-photon transition pathways results in EIA, which would strengthen the SWM process as well as the bipartite entanglement between the pump and probe fields with comparable intensity. Moreover, as analyzed in Ref. [8], the EIT term would decrease much more quickly than the TSTPE term with increasing the collisional decay rates due to the collisional decay rate $\gamma_{13p}$ equal to the sum of $\gamma_{12p}$ and $\gamma_{23p}$. This would lead to the change of the probe field absorption at the two-photon resonance from EIT to EIA, and subsequent enhancement of the bipartite entanglement with the increase of the collisional decay rates in a moderate range.

The above prediction is confirmed by solving the Heisenberg-Langevin equations and coupled propagation equations for the interaction of the pump and probe fields with atoms. The interaction Hamiltonian of the system in the rotating-wave approximation has the form [25-27],

$$\hat{V} = -\frac{\hbar N}{L}\int_0^L dz\left(\Delta_1\sigma_{22}(z,t)+(\Delta_1+\Delta_2)\sigma_{33}(z,t)+g_1 a_1(z,t)\sigma_{21}(z,t)+g_2 a_2(z,t)\sigma_{32}(z,t)+H.c.\right)$$

(2)

where $N$ is the total number of atoms in the interaction volume. The Heisenberg-Langevin equations and the coupled propagation equations are similar to those in Ref. [10], except that the two fields are applied in the counterpropagating configuration so as to eliminate the two-photon Doppler effect for the present case.

We use the similar analysis to that in Ref. [28] by writing each atomic or field operator as the sum of its mean value and a quantum fluctuation term to treat the interaction between the atoms and fields. We consider the case that $g_2\langle a_2\rangle$ is larger than $\sqrt{\gamma_{12}\gamma_{13}}$, so the depletions of the pump and probe fields can be safely neglected. To take into account the Doppler broadening, we assume the number of atoms per unit volume with velocity $v$ is $N(v)$ with the velocity distribution traditionally taken to be Maxwellian, and their contributions to the total atomic operators are obtained by integrating over the velocity distribution. The entanglement criterion $V_{12}=(\delta u)^2+(\delta v)^2<4$ proposed in Ref. [29] is employed to test the entanglement feature of the pump and probe fields, where $\delta u=\delta x_1-\delta x_2$ and $\delta v=\delta p_1+\delta p_2$ with $\delta x_i=(\delta a_i+\delta a_i^+)$ and $\delta p_i=-i(\delta a_i-\delta a_i^+)$ being the amplitude and phase quadrature fluctuation components of the quantum field operator $a_i$. Satisfying the above inequality sufficiently demonstrates the generation of bipartite entanglement, and the smaller the correlation $V_{12}$ is, the stronger the bipartite entanglement becomes. In the following, we assume the probe and pump fields to be initially in the coherent states $|\alpha_1\rangle$ and $|\alpha_2\rangle$, and the relevant parameters are scaled with m and MHz and set according to the realistic experimental conditions [30] with $r=4.5\times10^{-4}$, $L$=0.06, $\gamma_1$=3, $\gamma_2$=0.5, the atomic saturation density $n_0=8.5\times10^{15}$, and Doppler-broadened width $\Delta_w$=530 at room temperature.

Figure 2 gives the main result of this study, where the correlation $V_{12}$ at zero Fourier frequency and the probe absorption coefficient with the Doppler-broadening average at room temperature as a function of the probe field detuning $\Delta_1$ under different collisional decay rates $\gamma_{ijp}$ are depicted on the left column (A) and right column (B), respectively. It can be seen from the right column (B) that, similar to the results in Ref. [8], the line shape of the probe field absorption is changed from EIT to EIA with the increase of $\gamma_{ijp}$ in a moderate range. The most interesting thing is that on the left column (A), the evolution of the correlation $V_{12}$ almost exhibits an inverse behavior with increasing the collisional decay rates as compared to the probe absorption spectra. When there is no dephasing collisions (see p=0 in Fig. 2a), the correlation $V_{12}$ is always smaller than 4 in the whole Doppler-broadened range of the probe field detuning, which sufficiently demonstrates the generation of genuine bipartite entanglement between the pump and probe fields, and its line shape is a superposition of a sharp inverted dip with two narrow inverted peaks on its two sides superimposed on the inverted Doppler-broadened background, which results from the combination of EIT and TSTPE processes. However, the generated bipartite entanglement at the two-photon resonance is relatively weak due to the EIT effect. With the increase of $\gamma_{ijp}$, the line shape of the correlation $V_{12}$ would change from an narrow inverted dip into a distinct narrow inverted peak (see Fig. 2a-2c on the left column A). Further increasing $\gamma_{ijp}$ to the order of the pump field Rabi frequency would lead to a widely-broadened profile with a smaller reduction of $V_{12}$. This clearly indicates that the enhancement of the bipartite entanglement can be achieved via incoherent collisions in a moderate rang in the realistic inhomogeneously-broadened atomic system.

It is worthwhile to note that, one should not take it for granted that the stronger the nonlinear interaction with atoms accompanied by employing stronger light fields, the higher degree of the generated entanglement. This can be seen clearly from the dependences of the correlation $V_{12}$ on the pump field amplitude $\alpha_2$ for the cases p=0 and p=20 shown in Fig. 3a. In our calculations, in order to keep the mean values of atomic operators and intensity absorption rates of the two fields nearly stable so as to compare the degrees of entanglement under different field intensities, as done in Ref. [10], the ratios of $\alpha_2/\alpha_1$, $\gamma_{12,13,23}/\alpha_1$, and $n/\alpha_1$ are kept fixed. As seen in Fig. 3a, when the pump field is relatively weak, $V_{12}$ is almost equal to 4, so nearly no entanglement would exist between the probe and pump fields for both cases of p=0 and p=20. With the increase of $\alpha_2$, $V_{12}$ decreases gradually to be less than 4 for the case p=0, whereas it decreases rapidly to reach a minimum value of about 2.9 for the case p=20, which means that higher degree of entanglement can be achieved with larger intensities of the pump and probe fields and larger collisional decay rates in a moderate range. However, further increasing $\alpha_2$ would lead to the increase of $V_{12}$ and deterioration of the bipartite entanglement. For comparison, the probe absorption coefficient as a function of $\alpha_2$ is also shown in Fig. 3b. It is clear that the probe field absorption would decrease with increasing $\alpha_2$, that is, the EIT effect would be strengthened. This can be used to qualitatively explain the existence of an optimal pump field intensity for achieving the strongest entanglement. On one hand, the bipartite entanglement results from the nonlinear interaction between the atoms and laser fields through nonlinear multi-photon optical processes, which would be enhanced with increasing the pump and probe field intensities; on the other hand, as shown in Fig. 3b, the EIT effect would also be strengthened with the increase of the intensities of the two fields, which would trap the atoms in the level 1 and deteriorate the generation of entanglement. The optimal pump field intensity for the maximal degree of entanglement occurs when the effects of the two nonlinear processes balances each other. This can be further demonstrated by employing a stronger pump field ($\alpha_2 = 30\alpha_1$) with p=6 in Fig. 4(a-b). To compare with Fig. 2c and Fig. 2g, it can be seen that the narrow inverted entanglement peak in Fig. 2c turns into an inverted dip in Fig. 4a, whereas the narrow probe absorption peak in Fig. 2g turns into a dip in Fig. 4b. Obviously, the degree of bipartite entanglement would be dramatically reduced at the two-photon resonance due to the stronger EIT effect associated with the higher pump field intensity.

It is well-known that, in the EIT (EIA) scheme [8, 31, 32], the narrow dip (peak) due to destructive (constructive) interference takes place at the particular frequency position where the two-photon resonance condition is satisfied. This can also be confirmed in Fig. 4(c-d) with p=6 and the pump field detuning $\Delta_2$ equal to -200 for the present case. It is clear that the inverted narrow entanglement peak and narrow probe absorption peak move according to the pump field detuning and stand where the condition for the two-photon resonance is fulfilled, which indicates that the collision-induced entanglement peak does result from quantum interference between one-photon and multi-photon transition pathways.

Based on the above results, it can be inferred that the general thought that the incoherent collisions would always be detrimental to quantum entanglement is no longer true in spite of the dephasing nature, at least not in general. Moreover, the present scheme for generating and enhancing entanglement is conducted via incoherent collisions in the realistic inhomogeneously-broadened atomic system at room temperature or higher temperatures, which is very convenient for experimental implementation as compared to those using cold atoms [33-36].

In conclusion, we have demonstrated the enhancement of entanglement between two bright light fields via incoherent collisions in the traditional inhomogeneously-broadened ladder-type three-level atomic system. The strengthened entanglement results from the enhancement of constructive interference and suppression of destructive interference between one-photon and multi-photon transition pathways with the increase of the collisional decay rates in a moderate range. This method provides a promising alternative to generate and enhance nondegenerate continuous-variable entanglement between two bright light beams via incoherent collisions at room or higher temperatures, and may find potential applications in practical quantum information processing protocols.

## ACKNOWLEDGEMENTS

This work is supported by National Natural Science Foundation of China (Nos. 12174243). Yang's e-mail is yangxh@shu.edu.cn, and Xiao's e-mail is mxiao@uark.edu.## REFERENCES

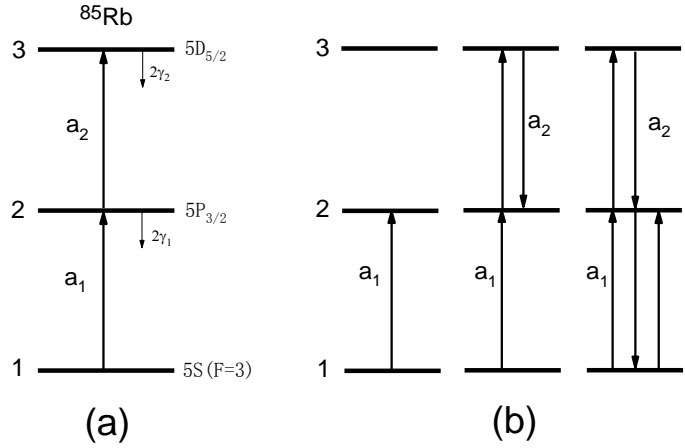

FIG. 1. (**a**) The ladder-type three-level atomic system driven by a strong coherent pump field ($a_2$) and a relatively weak probe field ($a_1$), where levels 1, 2, and 3 correspond, respectively, to the levels 5S (F=3), $5P_{3/2}$, and $5D_{5/2}$ of the $^{85}$Rb atom. (**b**) The one-photon, three-photon, and five-photon absorption up to the fifth-order terms of the probe field.

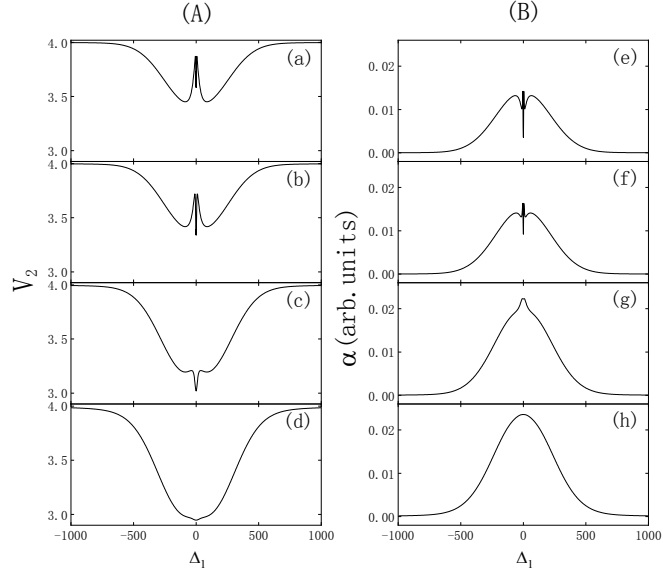

FIG. 2. The dependences of the correlation $V_{12}$ at zero Fourier frequency (the left column A) and the probe absorption coefficient (the right column B) with the Doppler-broadening average on the probe field detuning $\Delta_1$ with $r=4.5\times 10^{-4}$, $L$=0.06, $\gamma_1$=3, $\gamma_2$=0.5, $\alpha_2=5\alpha_1=50$, $n_0=8.5\times 10^{15}$, $\Delta_w$=530, $\gamma_{12p}=\gamma_{23p}$=1p and $\gamma_{13p}=\gamma_{12p}+\gamma_{23p}$=2p (p represents the relative collisional decay rate) under different collision-induced coherence decay rates p=0 (a, e), p=0.5 (b, f), p=6 (c, g), and p=20 (d, h), respectively.

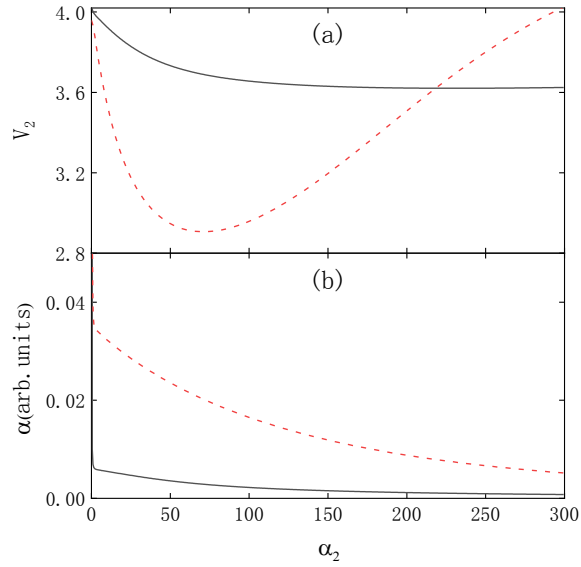

FIG. 3. (a) The dependences of the correlation $V_{12}$ at zero Fourier frequency (a) and the probe absorption coefficient (b) on the pump field amplitude $\alpha_2$ for the cases p=0 (solid black lines) and p=20 (dashed red lines), respectively, with $\alpha_2 = 5\alpha_1$, $n = n_0\alpha_1/10$, and $\gamma_{12,13,23}$ replaced by $\gamma_{12,13,23}\alpha_1/10$, and the other parameters are the same as those in Fig. 2.

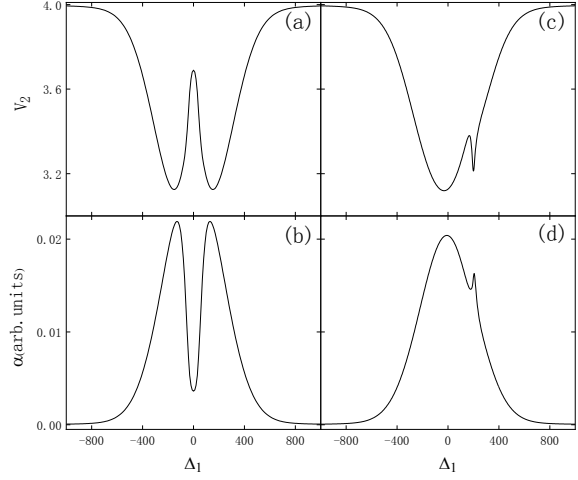

FIG. 4. The dependences of the correlation $V_{12}$ at zero Fourier frequency (a, c) and the probe absorption coefficient (b, d) on the probe field detuning $\Delta_1$ with p=6, $\alpha_2 = 30\alpha_1$ (a, b) and $\Delta_2 = -200$ (c, d), respectively, and the other parameters are the same as those in Fig. 2.